\documentclass[twocolumn,superscriptaddress,prl,amsmath,amstex,amssymb,citeautoscript]{revtex4-1}
\pdfoutput=1
\usepackage{natbib}
\usepackage[english]{babel}
\usepackage{letltxmacro}
\usepackage{latexsym}
\LetLtxMacro{\ORIGselectlanguage}{\selectlanguage}
\makeatletter
\DeclareRobustCommand{\selectlanguage}[1]{%
  \@ifundefined{alias@\string#1}
    {\ORIGselectlanguage{#1}}
    {\begingroup\edef\x{\endgroup
       \noexpand\ORIGselectlanguage{\@nameuse{alias@#1}}}\x}%
}
\newcommand{\definelanguagealias}[2]{%
  \@namedef{alias@#1}{#2}%
}
\makeatother
\definelanguagealias{en}{english}
\definelanguagealias{English}{english}
\usepackage{graphicx}
\usepackage{amsmath}
\usepackage{amsfonts}
\usepackage{amssymb}
\usepackage{bm}
\usepackage{color}
\usepackage[percent]{overpic}
\usepackage{soul} 
\usepackage{amssymb}
\usepackage{wasysym}
\usepackage{dsfont}
\usepackage{float}

\usepackage{hyperref}
\hypersetup{
    bookmarks=false,         
    unicode=false,          
    pdftoolbar=false,        
    pdfmenubar=true,        
    pdffitwindow=false,     
    pdfstartview={FitH},    
    pdftitle={},    
    pdfauthor={Authors},     
    pdfsubject={},   
    pdfcreator={},   
    pdfproducer={}, 
    pdfkeywords={many-body localization} {matrix elements} {disordered systems}, 
    pdfnewwindow=true,      
    colorlinks=true,       
    linkcolor=black,          
    citecolor=blue,        
    filecolor=magenta,      
    urlcolor=blue           
}

\newcommand{\be}{\begin{equation}}
\newcommand{\ee}{\end{equation}}
\newcommand{\bea}{\begin{eqnarray}}
\newcommand{\eea}{\end{eqnarray}}

\usepackage{graphicx}
\usepackage[colorinlistoftodos]{todonotes}
\usepackage{verbatim}

\newcommand{\ket}[1]{\mbox{$ | #1 \rangle $}}

\begin{document}

\title{Dynamically induced many-body localization}
\author{Soonwon Choi}
\affiliation{Department of Physics, Harvard University, Cambridge, Massachusetts 02138, USA}

\author{Dmitry A. Abanin}
\affiliation{Department of Theoretical Physics, University of Geneva, 1211 Geneva, Switzerland  }

\author{Mikhail D. Lukin}
\affiliation{Department of Physics, Harvard University, Cambridge, Massachusetts 02138, USA}

\begin{abstract}
We show that  a quantum phase transition from ergodic to many-body localized (MBL) phases can be induced via periodic pulsed manipulation of spin systems. Such a transition is enabled by the interplay between weak disorder and slow heating rates. Specifically, we demonstrate that  the Hamiltonian of a weakly disordered ergodic spin system can be effectively engineered, by using sufficiently fast coherent controls, to yield  a stable MBL phase, which in turn completely suppresses the energy absorption from external control field. Our results imply that a broad class of existing many-body systems can be used to probe non-equilibrium phases of matter for a long time, limited only by coupling to external environment.
\end{abstract}

\maketitle
Pulsed coherent manipulation is an indispensable tool in almost every branch of quantum science and technology.
First introduced as spin echo in nuclear magnetic resonance (NMR) experiments~\cite{Hahn_echo:1950ge}, a sequence of pulsed controls has proven highly successful in isolating quantum systems from unwanted noise sources.
Since then a variety of specialized dynamical decoupling techniques have been developed, ranging from frequency selective decouplings for quantum metrology to complex composite pulses for high fidelity quantum gate operations~\cite{deLange:2010ga,Lovchinsky:2016dq,Lovchinsky:2017gh,Brown:2004fs}.

Periodic manipulation of a many-body system has been utilized in order to effectively engineer interaction  and to probe exotic quantum phases of strongly interacting systems~\cite{Waugh:1968im,Lindner:2011ip,Rudner:2010bu,Iadecola:2015dg,Ponte:2015dc,Abanin:2016ev,Lazarides:2015jd,Khemani:2015gd,Else:2016gf,vonKeyserlingk:2016ev,Yao_dtc:2016wp}.
Indeed, in a number of systems ranging from ultracold atoms, molecules, and, ions to solid-state spin defects, coherent interactions among many particles and time-dependent controls are already being used for quantum simulations of  strongly correlated  dynamics~\cite{Yan:2013fna,Senko:2015hf,Zhang:2016uw,Choi:2016wn,Kucsko:2016tn,Wei:2016vp,Li:2016tc}.
Despite its apparent success, this pulsed Hamiltonian engineering approach is generally prone to heating and imprefections. 
In particular, it has been commonly believed that periodic external controls generically heat up an ergodic many-body system, eventually leading it to  infinite temperature, featureless  states~\cite{Lazarides:2014ie,DAlessio:2014fg,Ponte:2015vj}.
Likewise, any imperfections in pulsed manipulations may accumulate over a long time, resulting in uncontrolled dynamics.
Therefore, it may seem that the ultimate fate of any driven ergodic system corresponds to featureless, incoherent states.

This work demonstrates that a periodic control field can  \emph{induce} a phase transition of an isolated, ergodic system with weak disorder into a stable many-body localized (MBL) phase with completely suppressed energy absorption. In such a case, the system retains the memory of its initial state for asymptotically long time. In particular, these results also imply that dynamical decoupling with a finite repetition rate is sufficient for simulating MBL phases in existing experimental platforms for very long times  limited only by coupling to external environment.

This seemingly counter-intuitive phase transition can be understood as a consequence of the interplay between weak disorder  and parametrically suppressed heating~\cite{Abanin:2015bc,Mori:2016wb,Abanin:2015uh}.
Specifically, we focus on a situation where dynamical decoupling is employed to engineer an effective MBL hamiltonian that is  valid  for a long but finite lifetime $t^*$ without substantial heating.
We show that the resulting spectral properties of such a system further suppress energy absorptions, effectively extending $t^*$. Then, the evolution features 
completely suppressed heating, ultimately leading to the exact localization.
Furthermore, since localization  is robust against local perturbation, we find that the dynamically induced MBL  phase remains stable even in the presence of certain systematic experimental imperfections.

In what follows we first focus on a specific many-body spin model Hamiltonian, initially in the ergodic phase, and show that carefully chosen sequences of pulses can localize the system. 
We present analytical arguments illustrating the mechanism of the MBL transition as well as exact numerical simulations with finite size scaling supporting this conclusion. Finally, we generalize 
our analysis to a broad class of dynamical decoupling techniques.

\emph{Model.}---
We consider a chain of spin-1/2 particles with Heisenberg interactions between nearest neighboring pairs, 
described by the following Hamiltonian:
\begin{align}
\nonumber
H_0 &=\sum_i h_i S_i^z + \sum_i J\left( S_i^xS_{i+1}^x + S_i^y S_{i+1}^y + S_i^z S_{i+1}^z \right),
\end{align}
where $S_i^{\mu}$ ($\mu \in \{x,y,z\}$) is the Pauli spin-1/2 operator for a particle at site $i$, $h_i$ is a random on-site field uniformly and independently distributed among $[-W, W]$, and $J$ is the interaction strength between nearest neighboring spins. Dynamics governed by 
Hamiltonian $H_0$ has been explored in detail~\cite{Pal:2010gr,Luitz:2015tk}.  For a fixed value of $J$ the system is ergodic if the disorder strength $W$ is  smaller than a critical value $W_c$. For $W > W_c$, the system exhibits MBL dynamics. Extensive numerical simulations in Ref.~\cite{Pal:2010gr,Luitz:2015tk} suggest that $W_c/J \approx 3.5 \pm 1.0$. 

In what follows we focus on $W = J$, which resides deeply in the ergodic phase. However, the dynamics can be many-body localized 
by periodically applying pulses $P(\theta) = \exp{[-i \sum_{j}\theta S_{2j}^z]}$, which rotate every spin on even sites by an angle $\theta$ along the $\hat{z}$ axis. 
This conceptually simple sequence resembles a spin echo technique, which isolates the static magnetic field of a single spin from unwanted coupling to the environment. In our case,  it is used  to suppress spin exchange interactions while preserving on-site potential disorder.
When this pulse is repeated with period $\tau$, the system undergoes  dynamics governed by Floquet unitary~\footnote{We note that $U_F$ commutes with total magnetization along $\hat{z}$-axis $S_\textrm{total}^z \equiv \sum_i S_i^z$, making it easier to numerically simulate for a relatively large system sizes (up to $N=16$).
Below, we always work in the zero magnetization subspace.}:
\begin{align}
\label{eqn:floquet_unitary}
U_F(\theta, \tau) &= P(\theta)\exp{[ -i H_0 \tau]}.
\end{align}
The dynamics in Eq.~\eqref{eqn:floquet_unitary} can be understood by considering a time-\emph{dependent} Hamiltonian $H(t)$ defined in the so-called toggling frame~\cite{Waugh:1968im}.
As an example, for $\theta =\pi $ we work in the frame that rotates by $P(\pi)$ after each pulse. Similarly one can define a toggling frame for any angle $\theta = 2\pi (p/q)$ with integer $p,q \in \mathbb{Z}$ such that $H(t)$ is periodic in $q\tau$.
For $\theta = \pi$, the unitary evolution over two cycles can be written as 
\begin{align}
(U_F)^2 &= P(\pi) e^{-iH_0\tau} P(\pi) e^{-iH_0 \tau} \\
&= \mathcal{T} e^{-i \int_0^{2\tau} H(t) dt },
\end{align}
where we introduced a time-dependent Hamiltonian $H(t) = H_z + H_\perp (t)$ with
\begin{align}
H_z &= \sum_i h_i S_i^z + \sum_i J S_i^z S_{i+1}^z\\
H_\perp (t) &= \Theta (t) J \sum_i (S_i^xS_{i+1}^x + S_i^yS_{i+1}^y).
\end{align}
Here $H_z$ corresponds to the time averaged Hamiltonian of $H(t)$, and $H_\perp$ is the time-dependent component with rectangular envelope function $\Theta(t)$ that is periodic in $2\tau$: $\Theta (t) = \textrm{sgn}[\sin{(\pi t/\tau)}]$.

Since $H_z$ describes a trivial localized phase, 
our task is reduced to performing  the stability analysis of this phase upon the time-dependent driving $H_\perp (t)$. Such a problem has been analyzed in Ref.~\cite{Lazarides:2015jd,Ponte:2015dc,Abanin:2016ev,Burin:2017vz}, where it has been shown that a MBL system remains localized as long as the fundamental frequency $\omega_0$ of the time-dependent perturbation is large compared to perturbation strength $J$ and on-site disorder energy scale $h$. In our case,  $h \sim J$, and the required condition corresponds to a rapid repetition of the pulse sequence with $\omega_0 \equiv \pi / \tau \gg J$.
Therefore, these considerations indicate that  we can transform an ergodic system in to a MBL system via periodic pulse.

In order to clarify the mechanism of the localization, we next present  an intuitive picture of dynamically induced localization based on the combination of slow heating rates and spectral response of a typical MBL system. Underlying principles of the analysis are closely related to frameworks introduced in Ref.~\cite{Ponte:2015vj,Abanin:2015bc,Mori:2016wb,Abanin:2016ev}. 
We rewrite the envelope function as 
\begin{align}
\Theta (t) = \sum_m \frac{1-(-1)^m}{i\pi m} e^{i m\omega_0 t},
\end{align}
where $m$ enumerates harmonics of the fundamental frequency.
The components of this time-dependent perturbation become relevant only when they resonantly couple two many-body states with energy separation $\Delta E \approx m \omega_0$ for some $m \in \mathds{Z}$. When $\omega_0 \gg J $ and the perturbations are local, such resonant processes are absent since a rearrangement of a single spin alone cannot accommodate the absorption of a large energy quantum. Instead, $H_\perp (t)$ affects the dynamics of the system via higher order processes, which renormalize the effective Hamiltonian. These corrections can be perturbatively analyzed with a small parameter $J/\omega_0$, and it has been shown that, for a {\it generic} quantum many-body system, such perturbation theory is asymptotic with optimal order $k^* \equiv \omega_0 / J$~\cite{Abanin:2015bc,Mori:2016wb,Abanin:2016ev}. The physical meaning of $k^*$ is the minimal number of particles that need to cooperatively rearrange in order to absorb or emit one unit of energy quantum from external driving. The perturbative procedure integrates out the processes that affect $k<k^*$ spins, producing an effective Hamiltonian~\cite{Abanin:2015bc,Mori:2016wb,Abanin:2016ev}:
\begin{align}\label{eq:effective_H}
H_\textrm{eff} (t) = H_\textrm{eff}^*  + V^{*}(t), \,\,  H_\textrm{eff}^* =H_z + \sum_{k=2}^{k^*-1} H^{(k)}
\end{align}
where $H_\textrm{eff}^* $ denotes the static part of the effective Hamiltonian, which includes corrections $H^{(k)}$ up to order $k^*-1$, and $V^*$ contains all remaining time-dependent AC perturbations that can potentially heat up the system. While $V^*$ now contains $k\geq k^*$ spin processes, the quantum amplitude for rearranging $O(k)$ nearby spins becomes exponentially small in $k$:  $A_k \sim J (J/\omega_0)^{k-1}$. Hence, the dynamics of the system can be approximated by $H_\textrm{eff}^*$ for a long but finite time $t^* \sim 1/A_{k^*}$.

In our case, the static part of the effective Hamiltonian $H_\textrm{eff}^* $ is perturbatively close to $H_z$, and therefore it remains in the MBL phase for $J\ll \omega_0$.
Moreover, owing to this localization of $H_\textrm{eff}^*$ the remaining AC corrections given by $V^*(t)$ do not typically lead to resonant energy absorption, indicating that the system fails to heat up. This is in strong contrast to the case where the static part of the effective Hamiltonian describes an ergodic phase -- then the AC corrections do lead to resonant processes, and the system eventually heats up to infinite temperature (though at a rate exponentially slow in $\omega_0$)~\cite{Abanin:2015bc,Mori:2016wb,Abanin:2016ev}. 

To show that AC corrections are non-resonant, we consider $k>k^*$-th order perturbative process in $V^*(t)$.
Due to the locality of $H_\textrm{eff}^*$ and $H_\perp (t)$, such a process may rearrange only up to $k+2\xi$ spins, where $\xi$ is the localization length of $H_\textrm{eff}^*$.
Since typical many-body level spacing of $k$-spin rearrangement scales as $\delta_k \sim \sqrt{k} J / 2^{k}$, the probability of having a resonant $k$-body process becomes 
\begin{align}
P_k(\textrm{heating}) \sim\frac{ A_k}{ \delta_{k+2\xi}} \sim \frac{2^{2\xi+1}}{\sqrt{k+2\xi}} \left( \frac{ 2J}{\omega_0} \right)^{k-1},
\end{align}
which is exponentially small in increasing $k$ for $\omega_0 \gg J$. This indicates that the system fails to heat up and remains in the MBL phase.
\begin{figure}[tb]
\includegraphics[width=3.2in]{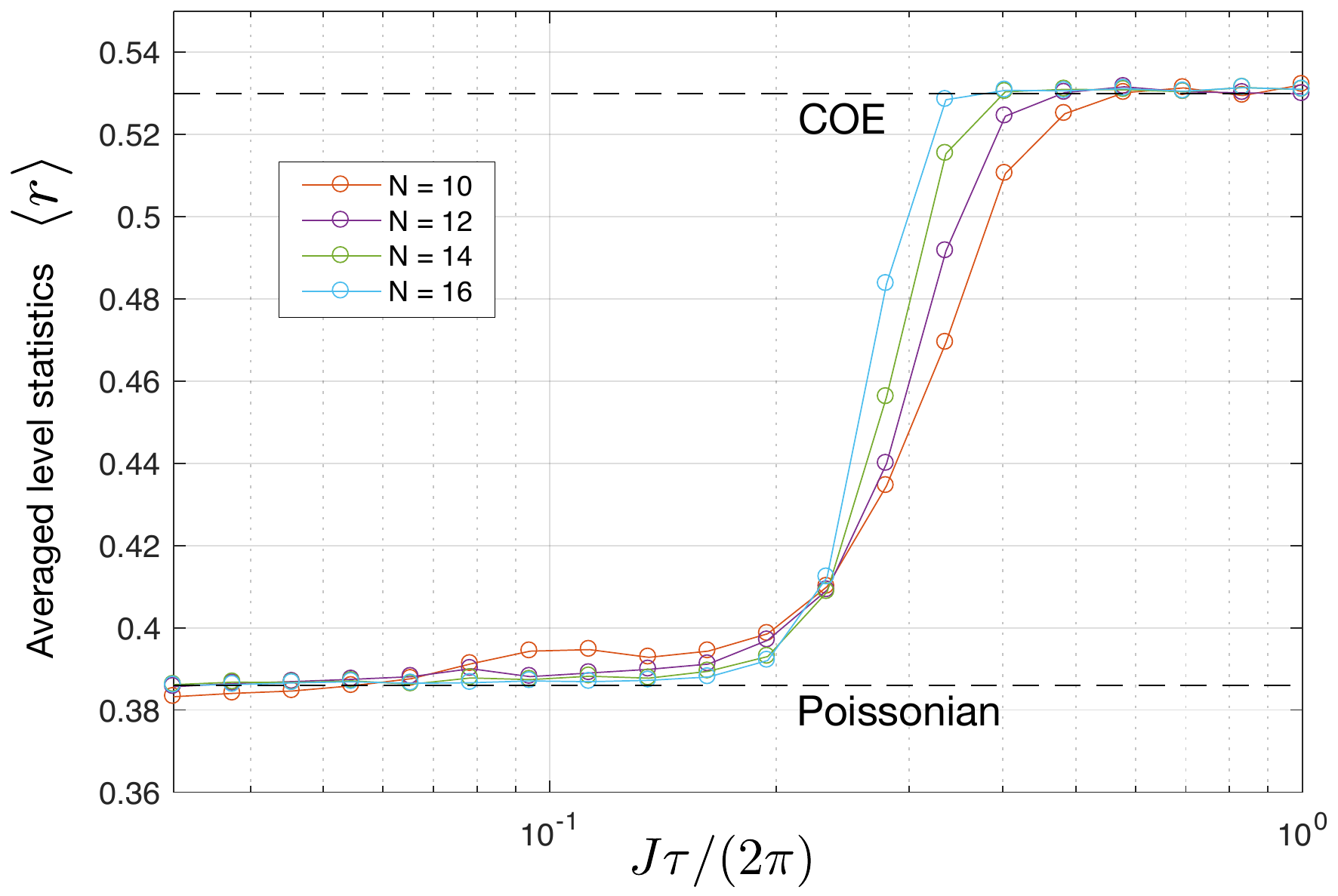}
\caption{Averaged level statistics $\langle r\rangle$ as a function of $\tau$ for various system sizes $N =10$, 12, 14, and 16.
Black dotted lines indicate the expected values of $\langle r \rangle$ in two limits:  the distribution from circular orthogonal ensemble (top) and Poissonian distributions (bottom).
For sufficiently fast pulses $\tau < \tau_c$, the level statistics approaches to the value corresponding to Poissonian distribution, indicating that the system belongs to a  MBL phase. The transition becomes sharper as system size increases. Each data point has been averaged over at least 100 disorder realizations } 
\label{fig_01}
\end{figure}

\emph{Numerical simulations.}---In order to corroborate our analytical arguments and check their self-consistency, we performed numerical simulations based on exact diagonalization of unitary evolution $U_F$ for system sizes up to  $N = 16$.
We extract quasi energy  $\epsilon_i \in [-\pi,\pi]$ from eigenvalues of $U_F$ by taking the imaginary parts of their logarithms.
We identify the MBL phase transition using a parameter $\langle r \rangle$ which characterizes level statistics of $\epsilon_i$: 
\begin{align}
\langle r \rangle = \left \langle \frac{\min{(\Delta \epsilon_i ,\Delta \epsilon_{i+1}) } }{\max{(\Delta \epsilon_i ,\Delta \epsilon_{i+1}) } } \right\rangle,
\end{align} 
where $\Delta \epsilon_i \equiv \epsilon_{i+1} - \epsilon_i$ and the averaging $\langle \cdot \rangle$ is taken over both the entire spectrum and disorder realizations of $U_F$.
If the system belongs to an ergodic phase $\langle r \rangle \approx 0.53$, corresponding to the value for a circular orthogonal ensemble (COE), while if it is in the MBL phase $\langle r \rangle \approx 0.386$, corresponding to the value for the Poisson statistics that lacks level repulsion.
We compute $\langle r \rangle$ as a function of $\tau$ and $\theta$ for varying system sizes $N = 10, \dots, 16$ as summarized in Fig.~\ref{fig_01}.
The value of $\langle r \rangle$ changes between two expected values as a function of $\tau$ (in units of $2\pi/J$). 
As the system size is increased, the transition of $\langle r \rangle$ values becomes sharper, suggesting a quantum phase transition in a thermodynamic limit. We extract a critical point $ 2\pi /\tau_c \sim 4J$. We note that, at this extracted critical point, the fundamental driving frequency $\omega_0$ is still smaller than the many-body band width $\sim 7J$ of the finite size system ($N=16$), confirming that our numerics cannot be explained by a trivial finite-size effect.

\begin{figure}[tb]
\includegraphics[width=3.2in]{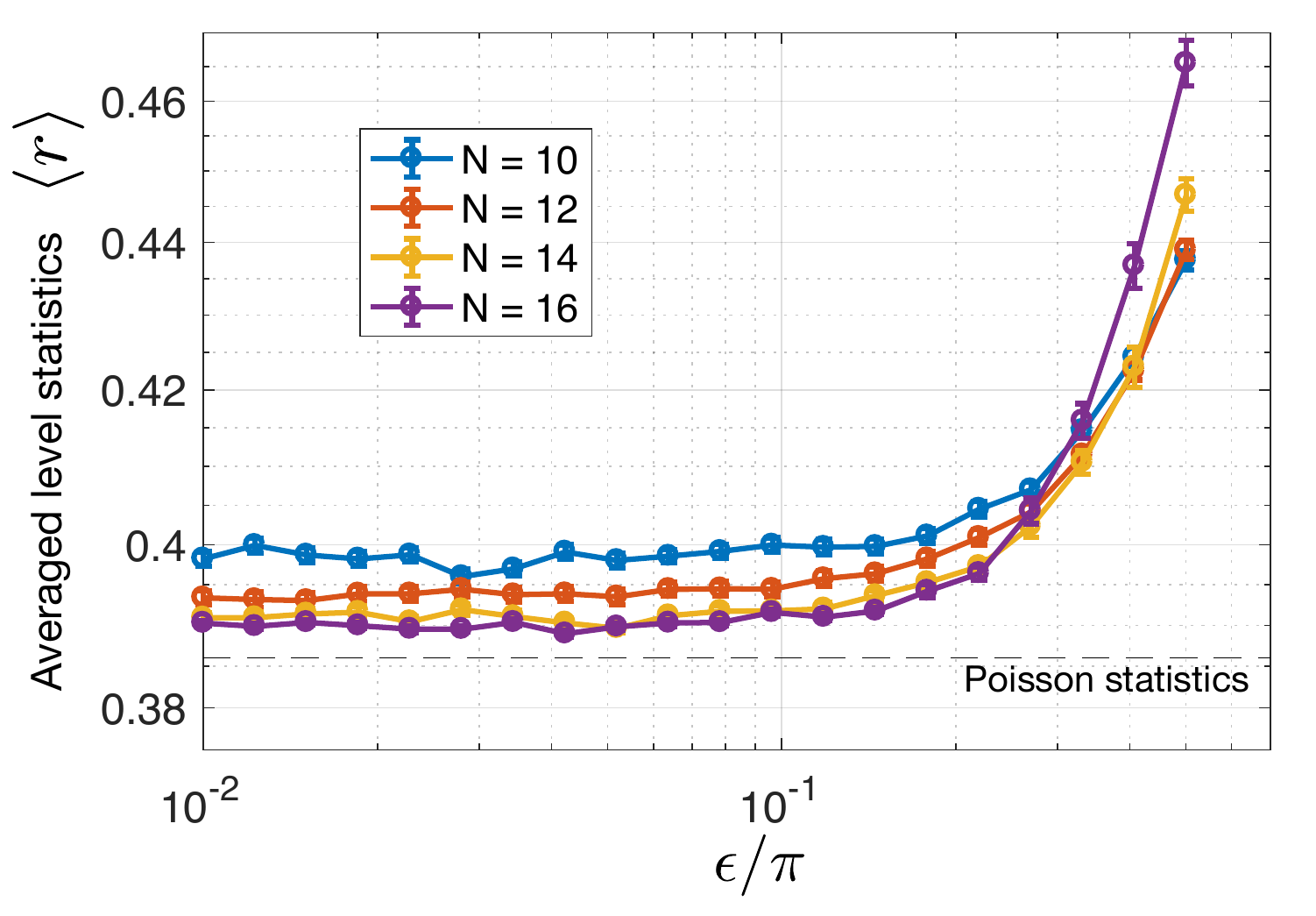}
\caption{Averaged level statistics $\langle r\rangle$ as a function of $\epsilon = \theta - \pi$ for a fixed value $J\tau =2\pi/ 5.5$. The finite size scaling suggest that the observed MBL transition is valid for a finite range of $\theta$ close to $\pi$. Each data point has been averaged over at least 100 disorder realizations.} 
\label{fig_03}
\end{figure}

In order to demonstrate the interacting nature of the MBL phase, we numerically probe the logarithmic growth of entanglement entropy~\cite{Bardarson:2012gc,Serbyn:2013he}.
For a system of size $N$, we prepare an initial state $\ket{\psi_0}$ with total $N/2$ spin up excitations such that every pair of nearby sites are oppositely polarized:
\begin{align}
\ket{\psi_0} = \prod_{i=1}^{N/2} \left[ \sqrt{2} (S_{2i-1}^x + S_{2i}^x)\right] \ket{\downarrow}^{\otimes N}.
\end{align}
\begin{figure}[tb]
\includegraphics[width=3.2in]{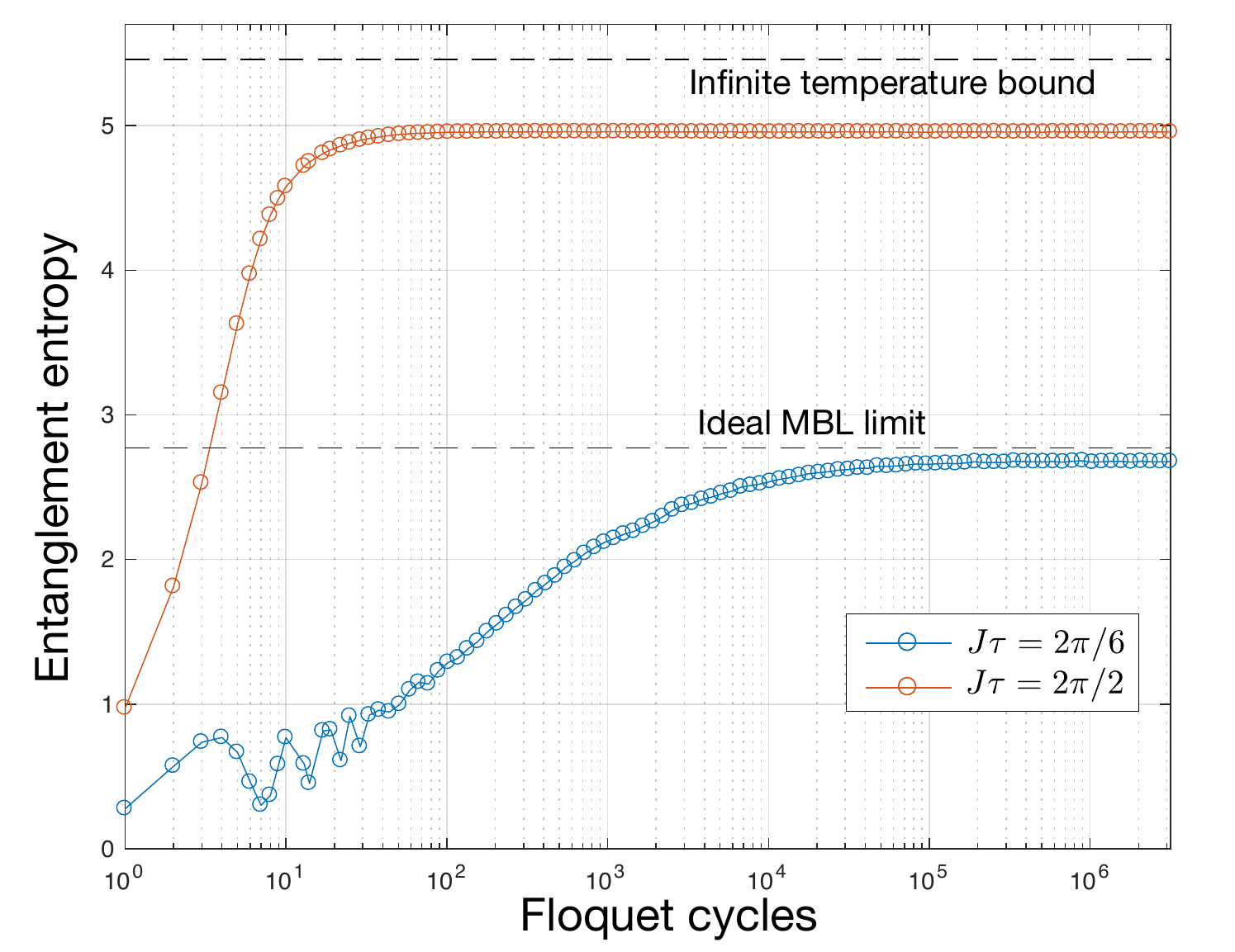}
\caption{Slow growth of entanglement entropy in a system of $N=16$ particles. For a sufficiently fast pulse sequence with $J\tau = 2\pi/6$ (blue), the entanglement entropy grows only logarithmically in time, while, for a slow pulse sequence with $J\tau =2\pi/2$ (red), it grows rapidly and saturates.
The saturation values are different in two cases since spin excitations cannot propagate in a localized phase.
Data has been averaged up to 100 disorder realizations. Two dotted lines indicate theoretical bounds for infinite temperature ensemble (top), in which all microscopic configurations are equally populated, and for an ideal MBL limit (bottom), in which spin excitations are completely localized while they still get entangled via Ising-type interactions.
}
\label{fig_02}
\end{figure}
After Floquet time evolution for $n$ cycles, we compute the entanglement entropy $S(n)$ along the cut at the middle of the system.
As illustrated in Fig.~\ref{fig_02}, we find qualitatively distinct behaviors in two cases: a long pulse period ($J\tau = 2\pi/2$) and a short pulse period ($J\tau = 2\pi/6$). In the former case, $S(n)$ quickly saturates to a value that is close to the theoretical bound $S_\infty$ of an infinite temperature ensemble. In the latter case, however, $S(n)$ grows logarithmically over multiple decades and saturates to a value $S_\textrm{MBL}$ that is much smaller than $ S_\infty$.
The difference between the two saturation values originates from the absence of transport in a localized phase, in which case entanglement entropy can only increase via phase correlations~\cite{Serbyn:2013he}.
Indeed, $S_\textrm{MBL}$ is close to the theoretical prediction corresponding to maximal entanglement entropy achievable from $\ket{\psi_0}$ for completely localized spin excitations but with phase correlations~\footnote{The theoretical bound in this case can be derived by treating each half of the chain as $N/4$ qubits made out of pairs of spins. The corresponding maximum entropy is given by $N/4 \log{(2)}$}.
Once saturated, the entanglement entropy does not grow further even for multiple decades, indicating the absence of slow heating.
We also check the robustness of our MBL phase with respect to finite deviation $\epsilon$ of the rotation angle $\theta$ from $\pi$.
As demonstrated in Fig.~\ref{fig_03}, we find that the MBL phase is stable over a range of $\epsilon$, in which a dynamical phase transition can be induced by a fast enough pulse sequence. 

\emph{Generalization and discussions}---
Our analysis can be generalized to a variety of dynamical decoupling techniques.
Consider, for example, a WAHUHA sequence that consists of four global spin rotations that are separated by uneven time durations~\cite{Waugh:1968im}. 
When the sequence is applied to a chain of dipolar interacting spin-1/2 particles with disordered on-site magnetic field,
the time averaged Hamiltonian displays exactly vanishing interactions while the disorder is only reduced by a constant factor, e.g. $h_i S_i^z \mapsto h_i (S_i^x - S_i^y + S_i^z) /3 $.
Therefore, as long as the total duration of the pulses is sufficiently short, the chain of spin-1/2 particles can be turned to a strict MBL system~\cite{Levitov:1990zz,Burin:2006ux,Yao:2014jj}, whose lifetime is only limited by coupling to the environment.
Indeed a closely related experiment has been already performed in Ref.~\cite{Wei:2016vp}, where slow development of correlations is observed in an effective one dimensional spin system.
More generally, we envision exploiting an ensemble of $d$-level systems with pair-wise short-range interactions and strong disorder. Such experimental settings are ubiquitous ranging from ultracold atoms, ions, molecules to solid state spin defects or superconducting qubits~\cite{Yan:2013fna,Senko:2015hf,Zhang:2016uw,Choi:2016wn,Kucsko:2016tn}.
One can design a finite $k$-pulse sequence with time separations $\tau_k$; if the sequence cancels the transport terms of the interactions and preserves weak disorder, one expects that a system can be dynamically induced to a MBL phase~\cite{DHEE2017}.

One intriguing  future possibility is to dynamically engineer Hamiltonians of long-range interacting systems~\cite{DHEE2017}.
On one hand, such a technique has already been used for observation of stable non-equilibrium states~\cite{Choi:2016wn} in the so-called critical regime~\cite{Kucsko:2016tn,CTC_abanin2017}, where the ergodicity is only marginally retained  via rare long-range resonances. On the other hand, recent work~\cite{Lee:2016dy} theoretically showed that the range of interactions can be effectively reduced via time modulated controls.
While the scheme presented in Ref.~\cite{Lee:2016dy} is relevant for short time evolution, the generalization of the scheme for asymptotically long time presents an intriguing avenue for future studies.
In combination with the present results, this may open the possibility of studying the interplay between long-range interactions and dimensionality of a system for a MBL phase transition, which still remains as an open question~\cite{Levitov:1990zz,Burin:2006ux,Yao:2014jj}.

We have demonstrated that an ergodic interacting system with weak disorder can be transformed into a MBL phase via dynamical decoupling techniques.
Our analytical arguments illustrate how the combination of slow heating and weak disorder leads to complete suppression of energy absorption.
From a practical perspective, our results provide a theoretical support for using driven systems for studying quantum phase transitions among MBL phases such as paramagnetic MBL to time-crystalline MBL.
Our results demonstrate that the non-equilibrium phases created in our approach can be stable against experimental imperfections and that their lifetimes are only limited by coupling to environment.
In addition, by tuning the pulse repetition rates, one can study the interplay between disorder and heating of a system.

\begin{acknowledgements}
We thank V.~Khemani, H.~Pichler, and W.~W.~Ho for useful discussions. This work was supported through NSF, CUA, the Vannevar Bush Faculty Fellowship, AFOSR Muri and Moore Foundation. S.~C. is supported by Kwanjeong Educational Foundation.
\end{acknowledgements}

\emph{Note added:} During the completion of this work, we became aware of a related contribution~\cite{Lindner2017dimbl}, in which driving induced MBL for ultracold atoms was suggested.

\bibliography{refs}

\end{document}